\documentclass[twocolumn]{aastex631}
\usepackage{amsmath}
\usepackage{graphicx}
\usepackage{xcolor}
\usepackage{dcolumn}
\usepackage{bm}
\usepackage{amssymb}
\usepackage{amsmath}
\usepackage{verbatim}
\usepackage[utf8]{inputenc}
\usepackage{tikz,hyperref}
\usepackage{dcolumn}

\shorttitle{SIDM Interpretations of Crater~II}

\shortauthors{Zhang, Yu, Yang, \& An}

\begin{document}

\title{Self-interacting dark matter interpretation of Crater~II}

\author[0009-0009-2791-1684]{Xingyu Zhang}
\email{zhang-xy19@mails.tsinghua.edu.cn}
\affiliation{Department of Physics, Tsinghua University, Beijing 100084, China}
\affiliation{Department of Physics and Astronomy, University of California, Riverside, California 92521, USA}

\author[0000-0002-8421-8597]{Hai-Bo Yu}
\email{haiboyu@ucr.edu}
\affiliation{Department of Physics and Astronomy, University of California, Riverside, California 92521, USA}

\author[0000-0002-5421-3138]{Daneng Yang }
\email{danengy@ucr.edu}
\affiliation{Department of Physics and Astronomy, University of California, Riverside, California 92521, USA}

\author[0000-0001-5431-6025]{Haipeng An}
\email{anhp@mail.tsinghua.edu.cn}
\affiliation{Department of Physics, Tsinghua University, Beijing 100084, China}
	\affiliation{Center for High Energy Physics, Tsinghua University, Beijing 100084, China}
	\affiliation{Center for High Energy Physics, Peking University, Beijing 100871, China}

\begin{abstract}

The satellite galaxy Crater~II of the Milky Way is extremely cold and exceptionally diffuse. These unusual properties are challenging to understand in the standard model of cold dark matter. We use controlled N-body simulations to investigate the formation of Crater~II in self-interacting dark matter (SIDM), where dark matter particles can scatter and thermalize. Taking the orbit motivated by the measurements from {\it Gaia} Early Data Release 3, we show a strong self-interacting cross section per particle mass of $60~{\rm cm^2/g}$ is favored for Crater~II. The simulated SIDM halo, with a $1~{\rm kpc}$ core, leads to both a low stellar velocity dispersion and a large half-light radius for Crater~II. These characteristics remain robust regardless of the initial stellar distribution.

\end{abstract}

\keywords{\href{http://astrothesaurus.org/uat/353}{Dark matter (353)};
\href{http://astrothesaurus.org/uat/420}{Dwarf spheroidal galaxies (420)};
\href{http://astrothesaurus.org/uat/1880}{Galaxy dark matter halos (1880)}}

\section{Introduction}
\label{sec:intro}

 Over the past decade, astronomical surveys have discovered more than $30$ satellite dwarf galaxies associated with the Milky Way, see, e.g.,~\cite{Simon:2019nxf}.  Among them, Crater~II, located $117~{\rm kpc}$ away from the Sun, is particularly interesting, as it has the coldest velocity dispersion of $\sigma_{\rm los} \approx2.7~{\rm km/s}$, while being exceptionally diffuse, with a projected half-light radius of $R_{1/2} \approx1.07~{\rm kpc}$~\citep{torrealba2016feeble-CraterII-discover,caldwell2017crater-CraterII-obs,Fu:2019,vivas2020decam,Ji:2021}. Crater~II has almost the lowest surface brightness among the satellites ever discovered. These unusual properties of Crater~II make it an intriguing case for testing the standard model of cold dark matter (CDM). 
 
Environmental effects can be important for the formation of Crater~II. The tides from its host Milky Way can strip away the halo mass and lower the velocity dispersion.~\cite{frings2017edge-tidalsim,applebaum2021ultrafaint-tidalsim} conducted cosmological hydrodynamical CDM simulations and argued that Crater~II-like galaxies could form through strong tidal interactions. However, more tailored studies found that it is difficult to simultaneously realize both low velocity dispersion and large size of Crater~II in CDM~\citep{sanders2018tidal-CraterII-arg,borukhovetskaya2022galactic-CraterII-sim}. The central cusp of a CDM halo is resilient to tidal stripping, and the tides must be sufficiently strong to explain the low velocity dispersion. Nevertheless, with such strong tides, the galaxy size would be truncated rapidly and it becomes much smaller than the observed size of Crater~II; see~\cite{borukhovetskaya2022galactic-CraterII-sim} for detailed discussion. It remains to be seen if baryonic feedback can modify these CDM predictions in accord with the observations of Crater~II.

In this work, we investigate the formation of Crater~II within the scenario of self-interacting dark matter (SIDM). Dark matter self-interactions transport heat and thermalize the inner halo; see~\cite{tulin2018dark-review,Adhikari:2022sbh} for reviews. For an SIDM halo in the core-expansion phase, a shallow density core forms and the stellar distribution correlates with the core size. A cored halo also boosts tidal mass loss and mild tides are allowed for lowering the halo mass. We conduct controlled N-body simulations to model the evolution of Crater~II in the tidal field of the Milky Way in SIDM, as well as in CDM for comparison. We will show that in SIDM the simulated galaxies can well reproduce the unusual properties of Crater~II. Taking the orbital parameters from {\it Gaia} EDR3~\citep{Gaia:2021,mcconnachie2020updated-pm-(5),Ji:2021,pace2022proper}, we will show that the required self-interacting cross section is $\sim60~{\rm cm^2/g}$ for Crater~II.

The rest of the paper is organized as follows: In Section~\ref{sec:setup}, we introduce our simulation setup. In Section~\ref{sec:results}, we discuss the properties of our simulated galaxies and compare them to the observations. We further discuss implications of our simulation results and conclude in Section~\ref{sec:conclusion}. In Appendix~\ref{sec:app1}, we discuss the choice of the final snapshot. In Appendix~\ref{sec:app2}, we show the degeneracy effect between tidal orbit and cross section. In Appendix~\ref{sec:app3}, we present additional N-body simulations with live stellar particles and further confirm the results discussed in the main text.

\section{Simulation setup}
\label{sec:setup}

In this section, we introduce initial conditions for modeling Crater~II in our N-body simulations, including initial halo and stellar density profiles and orbital parameters of our simulated satellite galaxies, the mass model of the Milky Way, and SIDM cross sections.

\subsection{The dark matter halo of Crater~II}
\label{sec:ic}

We consider three cases for the self-interacting cross section per particle mass: $\sigma/m=10~{\rm cm^2/g}$ (SIDM10), $30~{\rm cm^2/g}$ (SIDM30), and $60~{\rm cm^2/g}$ (SIDM60); as well as the CDM limit for comparison. For all cases, we model the initial halo of Crater~II with a Navarro-Frenk-White (NFW) profile~\citep{navarro1997universal-NFW2}
\begin{equation}
\label{eq-NFW}
\rho_{\mathrm{DM}}(r)=\frac{\rho_s}{\left(r / r_s\right)\left(1+r / r_s\right)^2},
\end{equation}
where $\rho_s$ and $r_s$ are the scale density and radius, respectively. For CDM, SIDM10, and SIDM30, we take $\rho_s = 1.42 \times 10^7~M_\odot/{\rm kpc}^3$ and $r_s = 2.06~{\rm kpc}$. The maximum circular velocity of the halo is $V_{\rm max}=26.57~{\rm km/s}$ and its associated radius is $r_{\rm max}=4.45~{\rm kpc}$. For SIDM60, $\rho_s = 1.14 \times 10^7~M_\odot/{\rm kpc}^3$ and $r_s = 2.23~{\rm kpc}$; $V_{\rm max}=25.77~{\rm km/s}$ and $r_{\rm max}=4.82~{\rm kpc}$. The halo parameters are overall consistent with those for progenitors of simulated satellite galaxies that have similar stellar masses of Crater~II~\citep{fattahi2018tidal-init-mass-Vmax}. As we will show later, with the tidal orbit from the measurements of {\it Gaia} EDR3~\citep{mcconnachie2020updated-pm-(5),Ji:2021,pace2022proper}, only SIDM60 can reproduce the observations of Crater~II. Thus we will focus on SIDM60 and CDM for the rest of the paper, and present SIDM10 and SIDM30 in Appendix~\ref{sec:app2} to show the degeneracy effect between orbit and cross section. 

There are stringent constraints on the cross section on the scales of galaxy clusters $\sigma/m\lesssim0.1~{\rm cm^2/g}$ for $V_{\rm max}\sim1500~{\rm km/s}$~\citep{kaplinghat2016dark-SIDM2,harvey2015nongravitational-SIDM-constraint,sagunski2021velocity-SIDM-constraint,andrade2022stringent-SIDM-constraint,rocha2013cosmological-SIDM-constraint,peter2013cosmological-SIDM-constraint} and elliptical galaxies~\citep{Kong:2024zyw}. Thus, a viable SIDM model must have a velocity-dependent cross section, which is strong in dwarf galaxies, while diminishing towards the clusters; see~\cite{tulin2018dark-review} and references therein for the discussion about particle physics realizations of SIDM. Nevertheless, we can use an effective constant cross section for a given halo to accurately characterize its gravothermal evolution~\citep{yang2022gravothermal-SIDM-example,Outmezguine:2022bhq,Yang:2022zkd,Yang:2023jwn,Fischer:2023lvl}. The constant cross sections we take should be regarded as effective cross sections for the Crater~II halo. 

We use the public code~\texttt {GADGET-2}~\citep{Springel:2000yr-GADGET2,Springel:2005mi-GADGET2}, which is implemented with an SIDM module developed and tested in~\cite{yang2022gravothermal-SIDM-example}, and the code~\texttt {SpherIC}~\citep{Garrison-Kimmel:2013yys} to generate the initial condition. For all cases, the initial halo mass is $3.37\times10^9~{M_\odot}$ and there are $10^7$ simulation particles. The particle mass is $337~M_\odot$ and the Plummer softening length is $7~{\rm pc}$. The resolution is high enough for the purpose of this work.

\subsection{The stellar component of Crater~II}
\label{sec:tagging}

Since Crater~II is dark matter-dominated, we can assume that stars are massless tracers of the halo potential and use the technique introduced in~\cite{bullock2005tracing-fE-method} to model the stellar component, i.e., the tagging method. In this work, we have further tested the method and validated its application in SIDM using N-body simulations with live stellar particles; see Appendix~\ref{sec:app3}. 

For each simulation particle, we can attach an appropriate probability to represent a star. The equilibrium distribution function of stars is~\citep{errani2020can-stellar-fE}
\begin{equation}
\label{eq:f}
f_\star(E)\equiv \frac{dN_\star}{d\Omega} =\frac{1}{\sqrt{8} \pi^2} \int_E^0 \frac{d^2 \nu_\star}{d \Phi ^2}\frac{d \Phi}{\sqrt{\Phi-E}}
\end{equation}
where $d\Omega = d^3r d^3 v$ is the differential volume of phase space, $\Phi$ the potential of the satellite halo, and $\nu_\star$ the stellar number density. It satisfies the normalization condition $\int dr ~ 4\pi r^2  \nu_\star(r) = N_\star$ with $N_\star$ being the total number of stars. We use an Einasto profile~\citep{einasto1965construction-Einasto} to model the stellar distribution of Crater~II
\begin{equation}
\rho_E(r)=\rho_E(0) \exp \left[-\left(\frac{r}{r_E}\right)^\alpha\right],
\label{eq:einasto}
\end{equation}
where $\rho_E (0)$ is the central density, $r_E$ is the scale radius, and we set the numerical factor $\alpha = 1$. For the Einasto profile, we have
\begin{equation}
\nu_\star(r) = \frac{N_\star}{8\pi r_{ E}^3}  {\rm e}^{-r/r_{ E}}.
\end{equation}
For each simulation particle located at $(\mathbf{r},\mathbf{v})$ in the phase space, the probability to represent a star is given by $P(E)\propto f_\star(E)/f_{\rm DM}(E)$, where $f_{\rm DM}(E)$ is the distribution function of dark matter particles. It is calculated in a similar way as in Equation~\ref{eq:f} with $\nu_\star$ being replaced by the number density profile of dark matter. We compute the probability at infall and reconstruct the stellar spatial distribution and kinematics from dark matter at later times by applying $P(E)$ as a weighting factor. 

\cite{torrealba2016feeble-CraterII-discover} originally used a Plummer profile to describe the stellar distribution and reported the 2D half-light radius as $R_{1/2}\approx1.066~{\rm kpc}$.~\cite{borukhovetskaya2022galactic-CraterII-sim} found that the Einasto profile in Equation~\ref{eq:einasto} ($\alpha=1$) fits the Plummer projected density profile well. For the purpose of comparison, we take the same approach and choose the initial Einasto scale radius to be $r_{E}=0.40~{\rm kpc}$, $0.73~{\rm kpc}$, and $1.37~{\rm kpc}$ as in~\cite{borukhovetskaya2022galactic-CraterII-sim}. Since the 2D half-light radius is $R_{1/2}=2.03 r_E$, we have the initial values $R_{1/2}\approx0.8~{\rm kpc}$, $1.5~{\rm kpc}$, and $2.8~{\rm kpc}$. These values bracket the measured value of $R_{1/2}\approx1~{\rm kpc}$, and they are overall consistent with those found for progenitors of satellite galaxies that have similar stellar masses of Crater~II in hydrodynamical CDM simulations~\citep{fattahi2018tidal-init-mass-Vmax}. 

For later snapshots, we fit the stellar distribution using the Einasto profile as well, extract the corresponding $r_E$ value, and calculate the half-light radius using the relation $R_{1/2}=2.03 r_E$. We have checked that using a Plummer profile in the tagging method gives rise to a similar evolution trajectory of the stellar distribution as with the Einasto profile. Furthermore, we have also confirmed that with the initial stellar distributions assumed in this work, the dark matter mass completely dominates over the stellar component in the entire evolution history of the simulated halo and the tagging method is well justified.

\subsection{The Milky Way}

We model the Milky Way with a static gravitational potential, which contains three components: a dark matter main halo, a stellar bulge, and stellar disks. The relevant parameters are given as follows.
\begin{itemize}
	\item 
	A spherical dark matter halo of an NFW profile
	\begin{equation}
	\Phi_{\rm halo} (r) = -4 \pi G \rho_s r_s^3 \frac{\ln \left(1+r/r_s\right)}{r},
	\end{equation}
	with $\rho_s = 7.7 \times 10^6~M_\odot/{\rm kpc}^3$ and $r_s = 20.2~{\rm kpc}$. The corresponding halo mass is $M_{200}\approx1.15\times10^{12}~M_\odot$ and the maximum halo circular velocity is $V_{\rm max}\approx192~{\rm km/s}$. 
	
	\item 
	A spherical bulge of a Hernquist profile~\citep{hernquist1990analytical}
	\begin{equation}
	\Phi_{\rm bulge} (r) = -\frac{G M_{ H}}{a_{H}} \frac{1}{1+r / a_{ H}},
	\end{equation}
	with $M_{H} = 2.1 \times 10^{10}~M_\odot$ and $a_{ H} = 1.3~{\rm kpc}$. 
	
	\item 
	Two axisymmetric disks of a Miyamoto-Nagai profile~\citep{miyamoto1975three}
	\begin{equation}
	\Phi_{\rm disk} (R,z) = -\frac{G M_{ d}}{\left[ R^2+\left(a_{ d}+\sqrt{z^2+b_{d}^2}\right)^2 \right]^{1/2}}.
	\end{equation}
	For the thin disk, we take $M_{d} = 5.9 \times 10^{10}~M_\odot$, $a_{ d} = 3.9~{\rm kpc}$, and $b_{d} = 0.3~{\rm kpc}$. For the thick disk $M_{ d} = 2 \times 10^{10} M_\odot$, $a_{d} = 4.4~{\rm kpc}$, and $b_{ d} = 0.92~{\rm kpc}$.
\end{itemize}

These values are the same as those in~\cite{borukhovetskaya2022galactic-CraterII-sim}, which are motivated by the Milky Way mass model in~\cite{mcmillan2011mass-MW} with a circular velocity of $239~{\rm km/s}$ at the solar radius $R_\odot = 8.3~{\rm kpc}$. We have checked that the pericenter of our simulated satellite halo would shift a few percent if taking the parameter values from the more recent model in~\cite{McMillan:2016jtx}. The main halo is treated as a static potential, and we neglect the evaporation effect between the satellite halo and the main halo~\citep{nadler2020signatures-core7,slone2023orbital-SIDM-example}. The approximation is justified if the cross section in the main halo is below ${\cal O}(1)~{\rm cm^2/g}$. This condition can be satisfied naturally for velocity-dependent SIDM models that evade the cluster constraints, see, e.g.,~\cite{yang2023strong-SIDM-example-model,Nadler:2023nrd} for examples.

\subsection{Orbital parameters}
\begin{table}[h]
	\centering
		\tabcolsep=0.25cm
	\begin{tabular}{lll}
		\hline
		$D_\odot$                                   &    $ 117.5 \pm 1.1 $                                &     \cite{torrealba2016feeble-CraterII-discover}      \\
		$v_{\rm los}$                               &    $ 87.5 \pm 0.4$                                   &     \cite{caldwell2017crater-CraterII-obs}      \\
		$\alpha_{\rm J2000}$                  &    $ 117.3^\circ $                                    &     \cite{torrealba2016feeble-CraterII-discover}      \\
		$\delta_{\rm J2000}$                   &    $ -18.4^\circ $                                     &     \cite{torrealba2016feeble-CraterII-discover}      \\
		$\mu_{\alpha^\star}$		&    $ - 0.07 \pm 0.02 $                            &     \cite{mcconnachie2020updated-pm-(5)}       \\
		$\mu_\delta$                		&    $ - 0.11 \pm 0.01 $                            &     \cite{mcconnachie2020updated-pm-(5)}       \\
		\hline
	\end{tabular}
	\caption{Measured orbital parameters of Crater II: the distance from the Sun $D_\odot $ [kpc], line-of-sight velocity in the solar rest frame $v_{\rm los}$ [km/s], right ascension $\alpha_{\rm J2000}$ and declination $\delta_{\rm J2000}$ [deg], proper motions $\mu_{\alpha^\star}$ and $\mu_\delta$ [mas/yr].}
	\label{tab:orbits1}
\end{table}

In Table~\ref{tab:orbits1}, we summarize the measured orbital parameters of Crater~II. Its proper motions are derived from data of the {\it Gaia} EDR3~\citep{mcconnachie2020updated-pm-(5),Ji:2021,pace2022proper}. In our simulation we have tested a series of obits with a wide range of pericenter $r_{\rm p}\approx2.5\textup{--}37.7~{\rm kpc}$. Our main results are based on two orbits, denoted as O2 and O3, as listed in Table~\ref{tab:orbits2}, where we have used the same naming scheme as in~\cite{borukhovetskaya2022galactic-CraterII-sim}. The orbit O2 has a pericenter of $r_{\rm p}\approx13.8~{\rm kpc}$ and it is {\it designed} such that the CDM halo can lose sufficient mass, while the orbit  O3 has $r_{\rm p}\approx37.7~{\rm kpc}$ and it is well consistent with the measurements of {\it Gaia} EDR3~\citep{mcconnachie2020updated-pm-(5),Ji:2021,pace2022proper}.

\begin{table}[h]
	\centering
	\tabcolsep=0.35cm
	\begin{tabular}{ c c c c c }
		\hline
		Orbit   &       $\mu_{\alpha^\star}$               &    $\mu_\delta$                   &    $r_{\rm p}$            &  Model  \\
		\hline
		O2      &                    -0.102                        &              -0.225                   &            13.8               &             CDM                \\
		O3      &                    -0.07                          &               -0.11                    &            37.7               &            CDM, SIDM60                  \\
		\hline   
	\end{tabular}
	\caption{Orbital parameters considered for our CDM and SIDM simulations. Columns
from left to right: Orbit name label, proper motions $\mu_{\alpha^\star}$ and $\mu_\delta$ [mas/yr], pericenter distance [kpc], and dark matter model. The orbit O3 is consistent with the measurements of {\it Gaia} EDR3~\citep{mcconnachie2020updated-pm-(5),Ji:2021, pace2022proper}, while O2 is designed for the CDM halo to lose sufficient mass.}
	\label{tab:orbits2}
\end{table}

\section{Results}
\label{sec:results}

In this section, we present the properties of our simulated galaxies and compare them with the observations of Crater~II. 

\subsection{Tidal evolution of the halo mass}
\label{section-mass-evolution}

\begin{figure*}[t]
		\centering
		\includegraphics[scale=0.33]{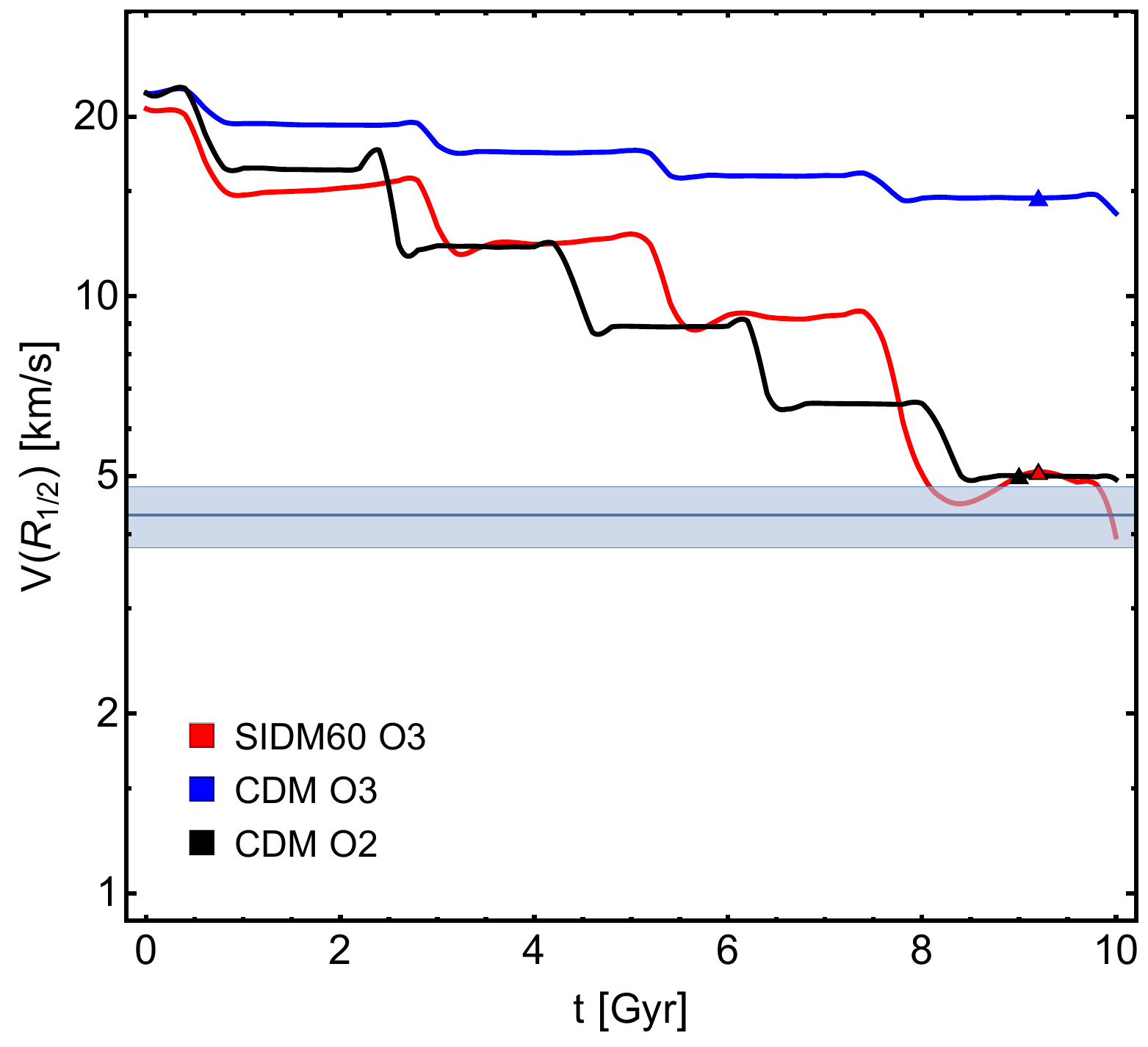}~~
		\includegraphics[scale=0.33]{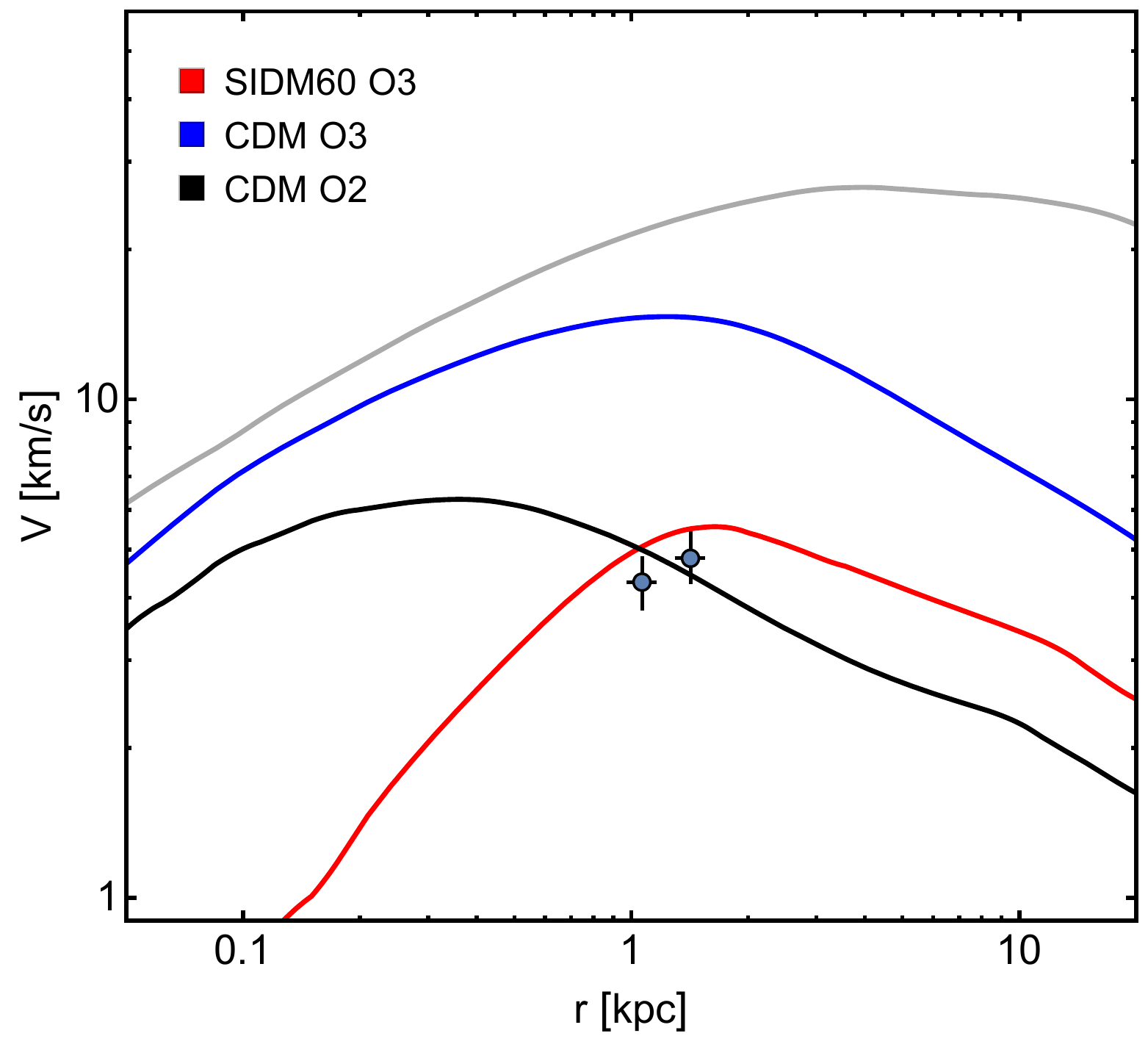}
	\caption{Left panel: Evolution of the circular velocity at half-light radius $V(R_{1/2}) \equiv \sqrt{G M(R_{1/2})/R_{1/2}}$ for SIDM60 O3 (red), CDM O3 (blue), and CDM O2 (black). The horizontal line and shaded region denote the measured value and its uncertainties $V(R_{1/2}) = 4.3 \pm 0.5~{\rm km/s}$~\citep{caldwell2017crater-CraterII-obs}. The triangle symbol denotes the final snapshot for comparing the simulated galaxies with Crater~II; they are relaxed after their last pericenter passage. Right panel: Corresponding circular velocity profiles at the final snapshot. The gray curve denotes the initial condition. The data points with error bars represent the measurements $V(R_{1/2})$ and $V(\frac{4}{3}R_{1/2})$ from~\cite{caldwell2017crater-CraterII-obs}. The orbit O3 is consistent with the measurements of Gaia EDR3~\citep{mcconnachie2020updated-pm-(5),Ji:2021, pace2022proper}, while the orbit O2 is designed such that the CDM halo has $V(R_{1/2})\approx5~{\rm km/s}$ at the final snapshot. 
}
\label{fig:vc}
\end{figure*}

In Figure~\ref{fig:vc} (left panel), we show the tidal evolution of the circular velocity at the half-light radius $V(R_{1/2})=\sqrt{GM(R_{1/2})/R_{1/2}}$, a proxy for the halo mass within $r=R_{1/2}=1.066~{\rm kpc}$, for the three cases: SIDM60 O3 (red), CDM O3 (blue), and CDM O2 (black). The horizontal line and shaded region denotes the measured value and its uncertainties $V(R_{1/2})=4.3\pm0.5~{\rm km/s}$~\citep{caldwell2017crater-CraterII-obs}. We choose the timescale marked by the triangle symbol as the final snapshot for analyzing other properties of the simulated galaxies, at which they are well relaxed. In Appendix~\ref{sec:app1}, we will show that our main results are robust to this choice. 

For SIDM60~O3, $V_{\rm} (R_{1/2})$ reaches the measured value after four pericenter passages. Thus, with the orbit motivated by the {\it Gaia} EDR3 measurements, an SIDM model with $\sigma/m\approx60~{\rm cm^2/g}$ is needed as the pericenter is large $r_{\rm p}=37.7~{\rm kpc}$. For CDM~O3, $V_{\rm} (R_{1/2})\approx15~{\rm km/s}$, a factor of $3$ larger than the measured value. If we choose the orbit O2, the CDM halo can reach $V(R_{1/2})\approx5~{\rm km/s}$ after five pericenter passages. However, its pericenter $r_{\rm p}\approx13.8~{\rm kpc}$ is a factor of $2.7$ smaller than that from the EDR3 measurements.

Figure~\ref{fig:vc} (right panel) shows the corresponding circular velocity profiles at the final snapshot for the three cases, as well as the initial condition. The measurements at $r=R_{1/2}$ and $r=\frac{4}{3}R_{1/2}$ are $V(R_{1/2})=4.3^{+0.5}_{-0.5}~{\rm km/s}$ and $V(\frac{4}{3}R_{1/2})=4.8^{+0.6}_{-0.5}~{\rm km/s}$~\citep{caldwell2017crater-CraterII-obs}, respectively. Tidal stripping can significantly reduce the overall halo mass, lowering the circular velocity in accord with the measurements at $r\sim R_{1/2}$ for SIDM60~O3 and CDM~O2. However, their $V(r)$ profiles are different, which can be understood as follows. For CDM~O2, $r_{\rm max}$ reduces to $0.4~{\rm kpc}$ at the final snapshot and the inner halo remains cusp. In the case of SIDM, the self-interactions push dark matter particles outwards. Thus, compared to CDM~O2, SIDM60~O3 has a lower inner density while larger $r_{\rm max}\approx1.5~{\rm kpc}$ at the final snapshot. We also see that the circular velocity of CDM~O3 is too high to be consistent with the measurements.

\subsection{The stellar distribution}

\begin{figure*}[h!]
		\centering
		\includegraphics[scale=0.325]{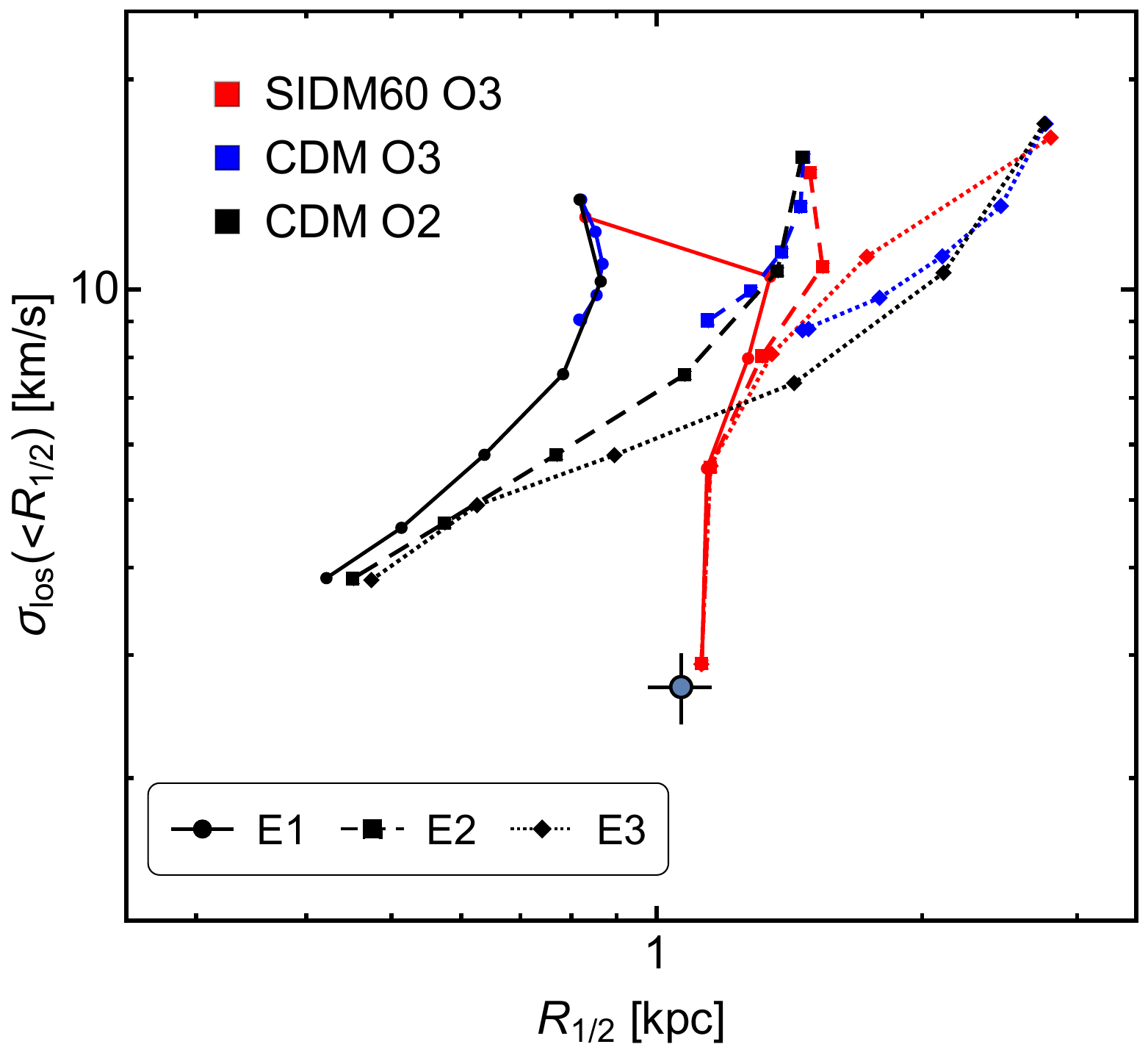}~~~
		\includegraphics[scale=0.325]{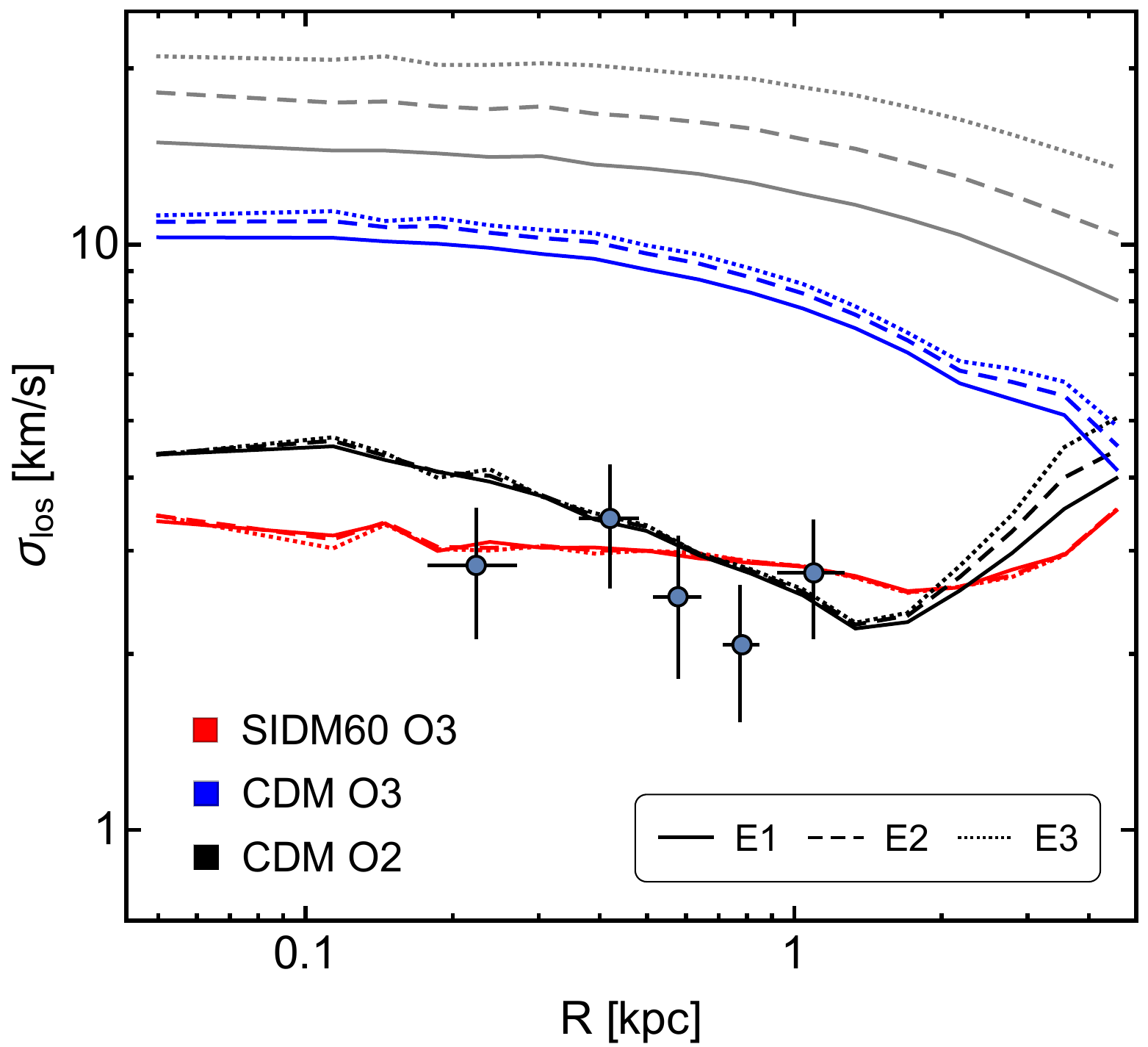}
		\caption{Left panel: the evolution of the stellar half-light radius $R_{1/2}$ and the line-of-sight velocity dispersion of stars within $R_{1/2}$ for SIDM60~O3 (red), CDM~O3 (blue), and CDM~O2 (black). The circle, square, and diamond symbols denote their corresponding snapshots after every pericenter passage for evaluation. Right panel: The line-of-sight velocity dispersion profile of stellar particles at the initial (gray) and final (colored) snapshots for SIDM60~O3 (red), CDM~O3 (blue), and CDM~O2 (black). For both panels, the stellar distribution follows an Einasto profile with three initial $R_{1/2}$ values $0.8~{\rm kpc}$ (E1; solid), $1.5~{\rm kpc}$ (E2; dashed), and $2.8~{\rm kpc}$ (E3; dotted); the data point with error bars are from the measurements of Crater~II~\citep{caldwell2017crater-CraterII-obs}. The orbit O3 is consistent with the measurements of {\it Gaia} EDR3~\citep{mcconnachie2020updated-pm-(5),Ji:2021, pace2022proper}, while the orbit O2 is designed such that the CDM halo can lose sufficient mass.}
		\label{fig:rh}
\end{figure*}

  \begin{figure*}[t]
			\centering
		\includegraphics[scale=0.33]{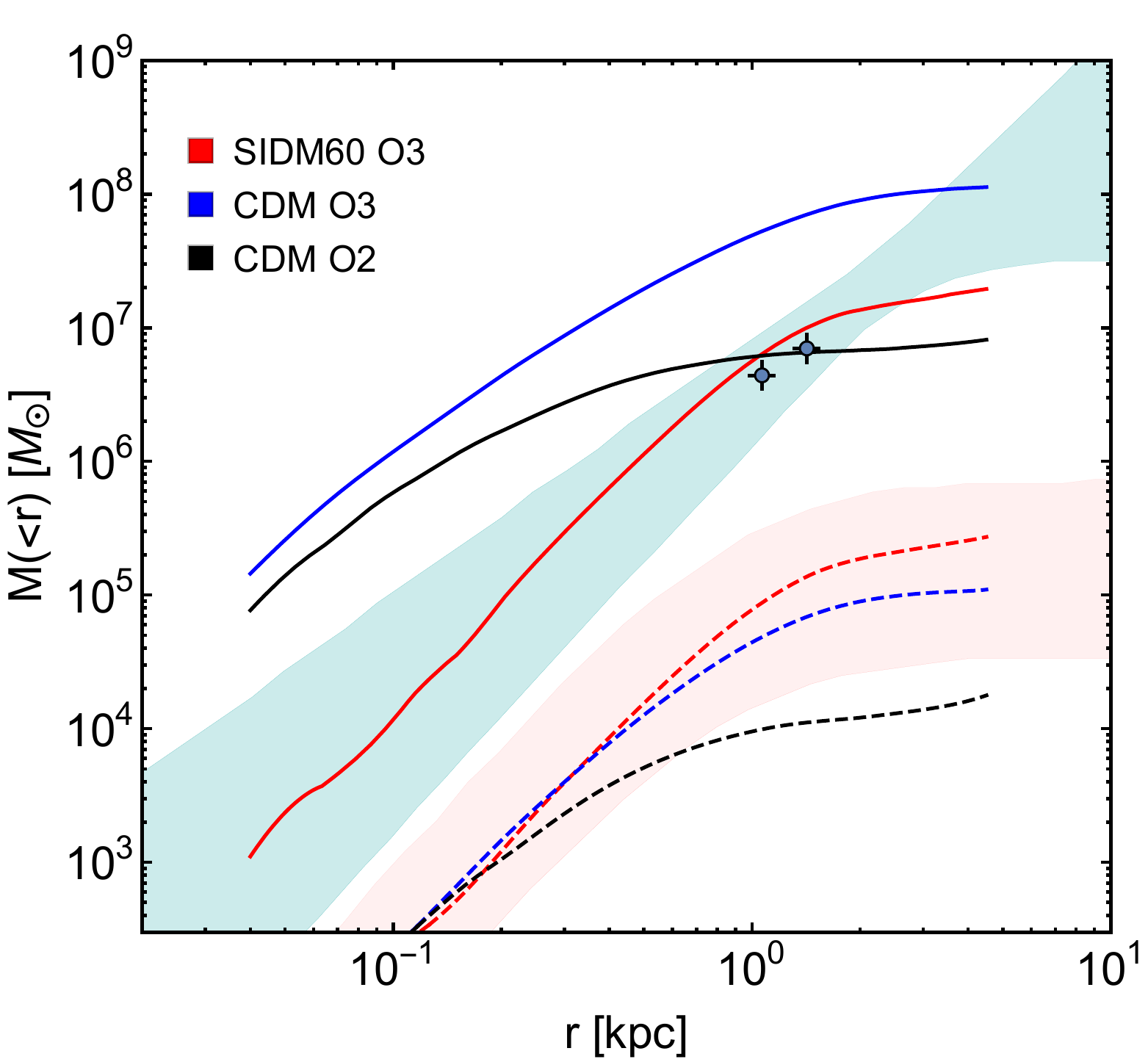 }~~~
		\includegraphics[scale=0.33]{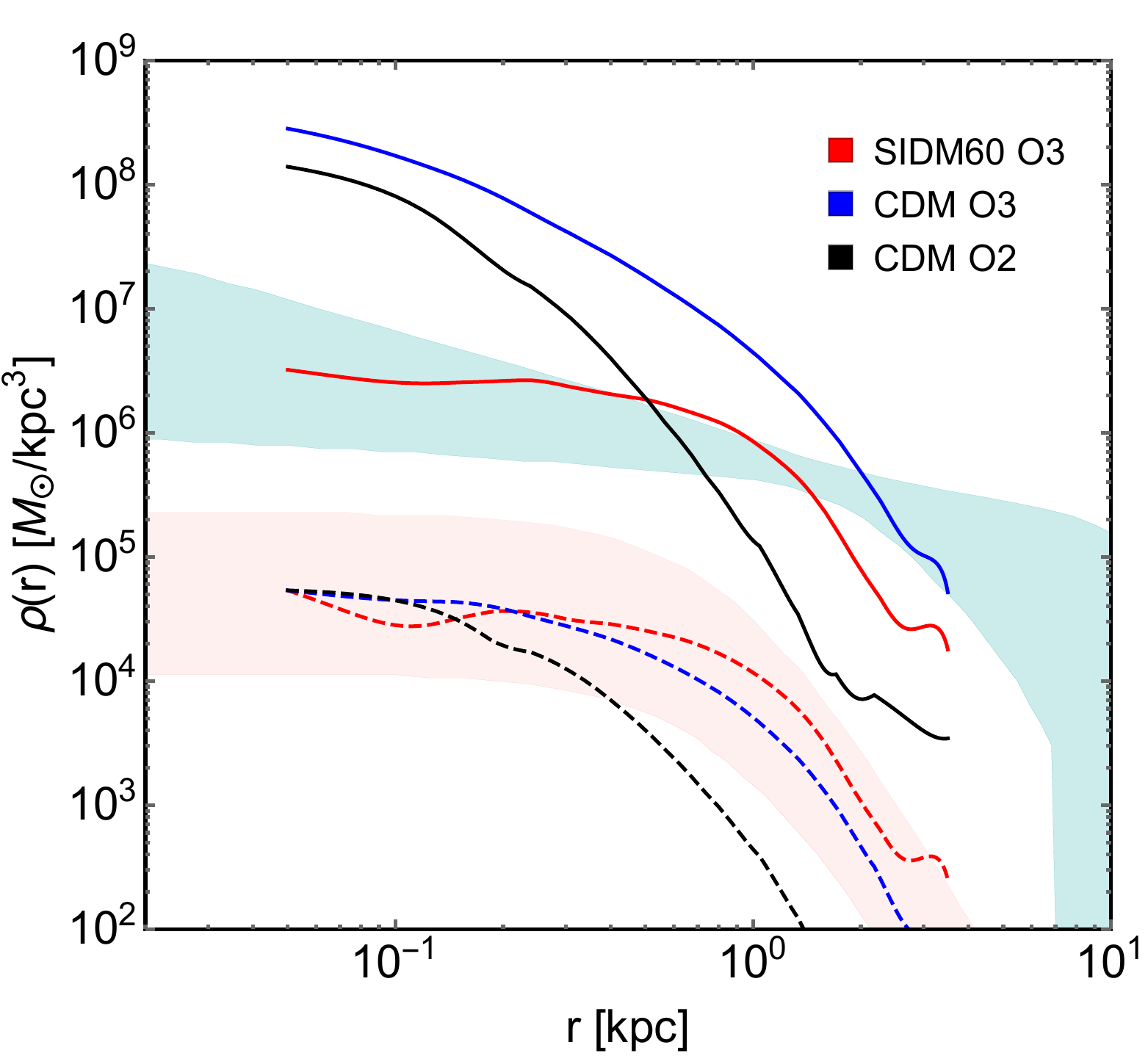}
	\caption{Mass (left panel) and density (right panel) profiles of the dark matter (solid) and stellar (dashed) components for SIDM60~O3 (red), CDM~O3 (blue), and CDM~O2 (black) at the final snapshot. The stellar density of the simulated galaxies is normalized to match the observed central stellar density of Crater~II.  The cyan and pink bands denote the $68\%$ credibility intervals for dark matter and stars, respectively, based on a fit to the kinematics of Crater~II~\citep{caldwell2017crater-CraterII-obs}. The data points in the left panel denote the enclosed masses evaluated within $R_{1/2}$ and $\frac{4}{3} R_{1/2}$~\citep{caldwell2017crater-CraterII-obs}.}
	\label{fig:density}
\end{figure*}

Figure~\ref{fig:rh} (left panel) shows the evolution of the half-light radius $R_{1/2}$ and the line-of-sight velocity dispersion $\sigma_{\rm los}$ of the stellar particles within $R_{1/2}$ for SIDM60~O3 (red), CDM~O3 (blue), and CDM~O2 (black). The initial conditions $R_{1/2}=2.03r_{E}\approx0.8~{\rm kpc}$, $1.5~{\rm kpc}$, and $2.8~{\rm kpc}$ are denoted as E1 (solid; dot), E2 (dashed; square), and E3 (dotted; diamond), respectively; for each case the symbols denote the evaluation at snapshots after every pericenter passage. 

For CDM~O2, the size converges to $R_{1/2}\approx0.3\textup{--}0.4~{\rm kpc}$ from $R_{1/2}\approx0.8\textup{--}2.8~{\rm kpc}$ after tidal evolution; see also~\cite{borukhovetskaya2022galactic-CraterII-sim}. We have checked that our final $R_{1/2}$ values agree with the estimates based on the tidal track of CDM satellite halos in~\cite{Errani:2021rzi}. The same trend applies to CDM~O3, but the convergence is slower as it has weaker tides. It is clear that both CDM O2 and O3 cannot explain the observations of Crater~II. The former's stellar size is a factor $3$ smaller than Crater~II's $R_{1/2}\approx1~{\rm kpc}$, while the latter's velocity dispersion is a factor of $3$ higher than the measured one $\sigma_{\rm los}(<R_{1/2})\approx2.7~{\rm km/s}$. Remarkably, SIDM60~O3 can simultaneously explain both stellar size and velocity dispersion of Crater~II. 

The upper limit of the galaxy size is largely set by $r_{\rm max}$. For the region beyond $r_{\rm max}$, stellar particles are stripped away significantly. From Figure~\ref{fig:vc} (right panel), we see that $r_{\rm max}\sim0.4~{\rm kpc}$ for CDM~O2, while $r_{\rm max}\sim1.5~{\rm kpc}$ for SIDM60~O3 at the final snapshot. Thus the latter can retain a larger galaxy size. More specifically, in SIDM there are two competing effects in affecting the size of the satellite galaxy: tidal truncation and SIDM core formation. The former tends to reduce the size, while the latter causes it to expand~\citep{Vogelsberger:2014pda,Carleton:2019,yang2020self-SIDM-UDG2}, see also Appendix~\ref{sec:app3}. If the initial distribution of stars is diffuse ($R_{1/2}\approx2.8~{\rm kpc}$), tidal truncation dominates and the final size is reduced. If it is compact ($R_{1/2}\approx0.8~{\rm kpc}$), the halo core expansion drives the increase of the stellar size. For $\sigma/m\approx 60~{\rm cm^2/g}$, SIDM produces a halo core size of $1~{\rm kpc}$, which is comparable to $R_{1/2}$ of Crater~II. We have also checked that the tidal radius of the SIDM60 halo is $1~{\rm kpc}$ at the final snapshot.

In Figure~\ref{fig:rh} (right panel), we show the line-of-sight velocity dispersion profile of stellar particles at the initial (gray) and final snapshot for SIDM60~O3 (red), CDM~O3 (blue), and CDM~O2 (black). For each of the three cases, despite the difference among three initial $\sigma_{\rm los}$ profiles corresponding to E1 (solid), E2 (dashed) and E3 (dotted), the final $\sigma_{\rm los}$ profiles largely converge. For SIDM60, the profile is consistent with the measurements~\citep{caldwell2017crater-CraterII-obs}. It is almost flat for $R\lesssim1~{\rm kpc}$ because the halo has a shallow density core, while $\sigma_{\rm los}$ increases in the region for $R\gtrsim2~{\rm kpc}$, where more stars are escaping at large radii. For CDM~O2, $\sigma_{\rm los}$ increases towards the center as expected from a central density cusp. For CDM~O3, $\sigma_{\rm los}$ is too high to be consistent with the measurements. We have further checked the $\sigma_{\rm los}$ profile using data from~\cite{Ji:2021} and found it overall agrees with that from~\cite{caldwell2017crater-CraterII-obs}.

\subsection{The mass and density profiles}

Figure~\ref{fig:density} shows mass (left panel) and density (right panel) profiles of the dark matter (solid) and stellar (dashed) components for the three cases at the final snapshot. For comparison, we also display the favored ranges of Crater~II from~\cite{caldwell2017crater-CraterII-obs} (cyan: halo; pink: stars; $68\%$ credibility intervals), based on a fit using a generalized NFW profile for the halo, allowing a density core. We see that SIDM60 reproduces both halo and stellar profiles inferred from Crater~II. This is not the case for the CDM halos.

\section{Discussion and Conclusions}

\label{sec:conclusion}

As we have shown, the exceptionally large size of Crater~II implies that its halo has a $1~{\rm kpc}$ density core. In CDM, a core could form due to baryonic feedback~\citep{read2005mass-bf,mashchenko2008stellar-bf,pontzen2012supernova-bf,di2014dependence-bf,chan2015impact-bf}. However, for a progenitor halo with mass $3\times10^9~M_\odot$, similar to Crater~II, the FIRE2 simulation shows the core size is ${\cal O}(10)~{\rm pc}$~\citep{Lazar:2020pjs}, too small to play a role. More work is needed to test if baryonic feedback can help reconcile CDM with the observations of Crater~II. 

Our SIDM interpretation of Crater~II can be extended to Antlia~II, another satellite with a large size $R_{1/2}\sim2.9~{\rm kpc}$ and $\sigma_{\rm los}\approx5.7~{\rm km/s}$~\citep{Torrealba:2018fwy}, challenging CDM as well~\citep{Errani:2021rzi}.~\cite{Ji:2021} found that there is a velocity gradient in Antlia~II and a tentative one in Crater~II. This may provide another test on the SIDM interpretation. Furthermore, there are over $60$ confirmed or candidate satellite galaxies of the Milky Way, see, e.g.,~\cite{mcconnachie2012observed-MW-satellite,drlica2020milky-MW-satellite}. They exhibit a great diversity in both size and enclosed mass, with $R_{1/2}$ ranging from $\sim10~{\rm pc}$ to a few ${\rm kpc}$ and $V(R_{1/2}) \sim 4$ to more than $10~{\rm km/s}$. Thus it is intriguing to test if SIDM can explain the full range of the diversity in the $V(R_{1/2})\textup{--}R_{1/2}$ plane of the satellite galaxies.   

Cosmological N-body simulations show that strong dark matter self-interactions can diversify inner dark matter densities of satellite halos~\citep{zavala2019diverse-SIDM-example,turner2021onset-SIDM-example,Correa:2022dey,yang2023strong-SIDM-example-model,Nadler:2023nrd}, due to core formation and collapse~\citep{balberg2002self-collapse1,koda2011gravothermal-collapse3}. For $\sigma/m\gtrsim30\textup{--}50~{\rm cm^2/g}$ in ultra-faint dwarf halos, the collapse could occur within the Hubble time, yielding a high central density. The collapse timescale is extremely sensitive to the concentration $\propto (\sigma/m)^{-1} c^{-7/2}_{200}$~\citep{essig2019constraining-tc,Kaplinghat:2019svz,Nadler:2023nrd,Zeng:2021ldo,Zeng:2023fnj}. The concentration of the Crater~II halo is close to the median and hence it is in the core-expansion phase. Nevertheless, the dense satellites of the Milky Way, such as Draco, could be in the collapse phase~\citep{nishikawa2020accelerated-collapse4,Sameie:2019zfo,kahlhoefer2019diversity-UDG1,Correa:2020qam}. Interestingly, $\sigma/m\sim60~{\rm cm^2/g}$ on the Crater~II mass scale is aligned with SIDM models proposed to explain the extreme diversity of dark matter distributions in other galactic systems, see, e.g.,~\cite{yang2023strong-SIDM-example-model,Nadler:2023nrd}. For the velocity-dependent SIDM model in~\cite{yang2023strong-SIDM-example-model}, the effective cross section on average is $\sim60~{\rm cm^2/g}$ at $V_{\rm max}\sim5\textup{--}25~{\rm km/s}$, the relevant range for the Crater~II halo.

In summary, we have performed controlled N-body simulations and shown that the unusual properties of Crater~II can be explained in SIDM. Dark matter self-interactions lead to core formation and boost tidal mass loss for the satellite halo, resulting in a low velocity dispersion of stars even if the halo has a relatively large pericenter. At the same time, the stellar distribution expands as the halo core forms and the stellar distribution correlates with the core size. In the future, we will expand our study to other satellite galaxies of the Milky Way and explore SIDM models that can explain the full range of the diversity of the satellites, as well as galaxies in the field~\citep{Gentile:2004tb,KuziodeNaray:2009oon,Oman:2015xda,Ren:2018jpt,PinaMancera:2021wpc,Kong:2022oyk,Montes:2023ahn,ManceraPina:2024ybj}.

\section*{acknowledgments}
	We thank Ethan Nadler and Laura Sales for comments on the manuscript. This work of H.-B.Y. and D.Y. was supported by the John Templeton Foundation under grant ID\#61884 and the U.S. Department of Energy under grant No.~de-sc0008541. The work of H.A. was supported in part by the National Key R\&D Program of China under Grants No.~2023YFA1607104, 2021YFC2203100, and 2017YFA0402204, the NSFC under Grant No.~11975134, and the Tsinghua University Dushi Program. The opinions expressed in this publication are those of the authors and do not necessarily reflect the views of the funding agencies.

\appendix

\section{Choice of the final snapshot}
\label{sec:app1}

\begin{figure}[h]
	\centering
	\includegraphics[scale=0.33]{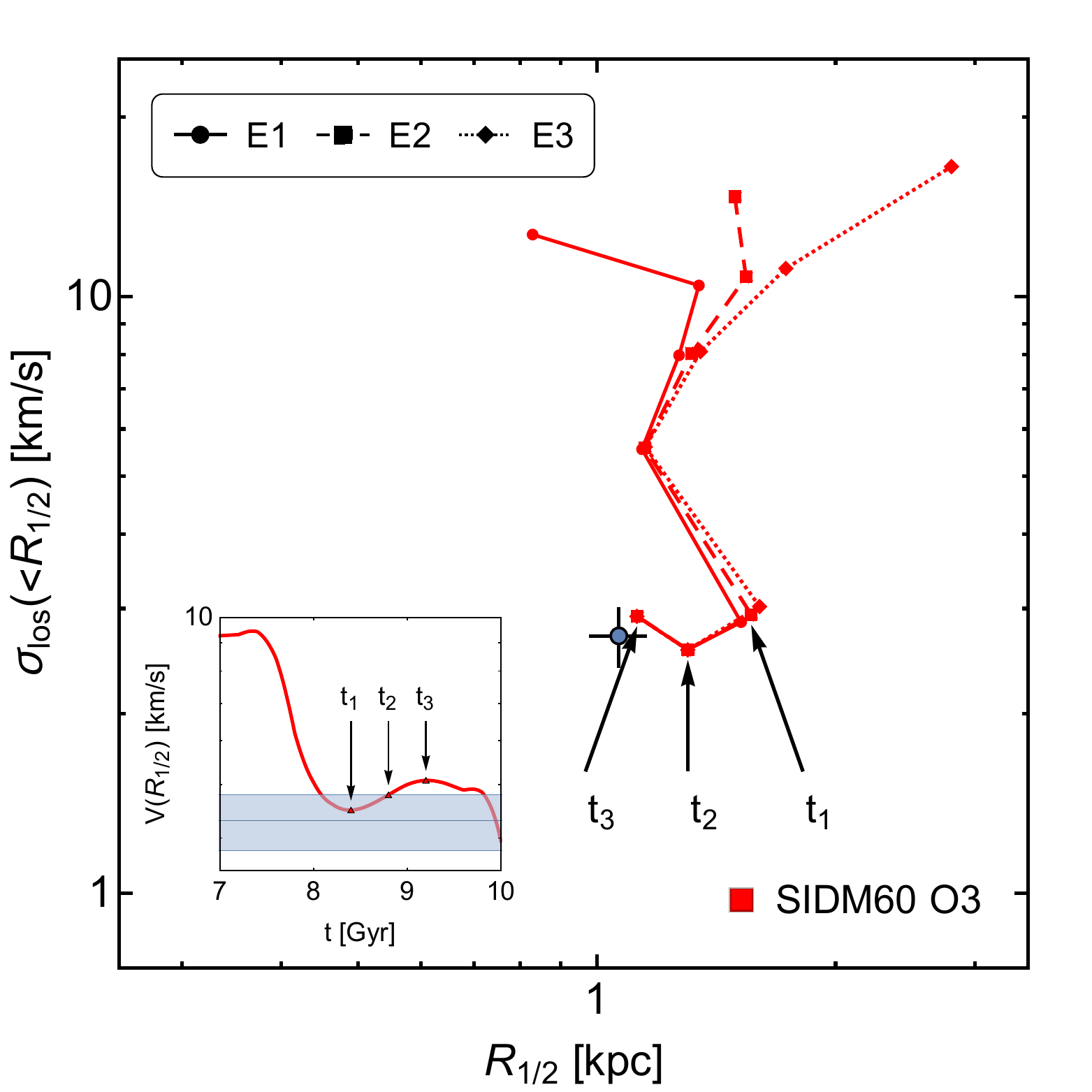}~~
	\includegraphics[scale=0.3155]{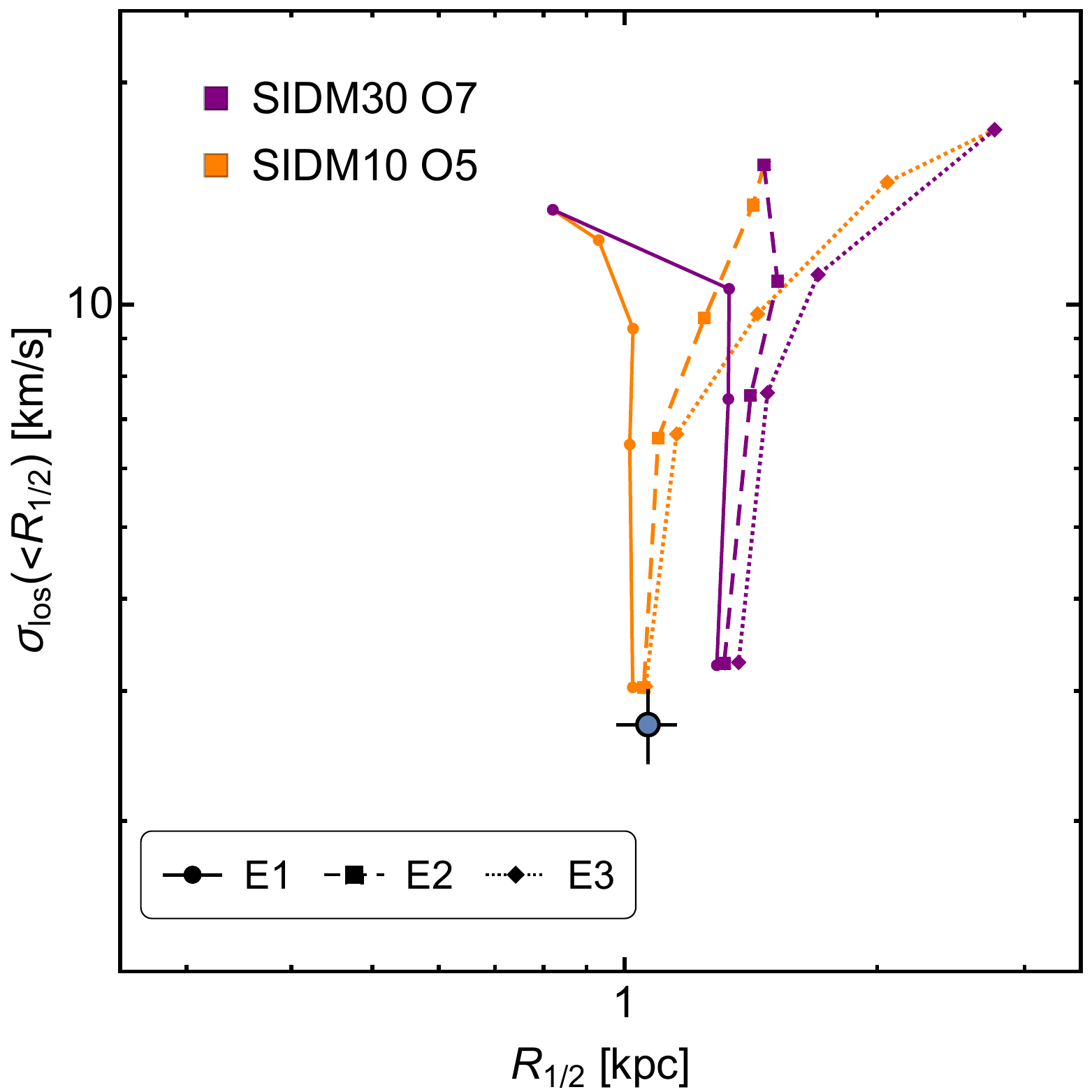}
	\caption{Left panel: Similar to the left panel of Figure~\ref{fig:rh}, but with three sequential final snapshots for SIDM60~O3 denoted as $t_1$, $t_2$, and $t_3$. The insert panel shows their corresponding locations in the $V(R_{1/2})\textup{--}t$ plane. Right panel: Similar to the left panel of Figure~\ref{fig:rh}, but for two additional SIDM simulations SIDM10~O5 (orange) and SIDM30~O7 (purple). 
}
\label{fig:app}
\end{figure}

Figure~\ref{fig:app} (left panel) shows the $R_{1/2}$ and $\sigma_{\rm los}(<R_{1/2})$ values for SIDM60~O3 with three different final snapshots denoted as $t_1$, $t_2$, and $t_3$. For our main results, we showed the properties of the simulated galaxies at the snapshot when they are well relaxed, i.e., $t_3$ for SIDM60. The results remain largely the same if we evaluate the properties at the three different final snapshots.  

\section{Degeneracy}
\label{sec:app2}

Figure~\ref{fig:app} (right panel) shows the evolution of $\sigma_{\rm los}(<R_{1/2})\textup{--}R_{1/2}$ for two additional SIDM simulations SIDM10~O5 (blue) and SIDM30~O7 (magenta), where the parameters of the orbit O5 (O7) are $\mu_{\alpha^*}=-0.091~(-0.0807)$, $\mu_\delta=-0.187~(-0.148)$, and $r_{\rm p}\approx19.9~(27.9)~{\rm kpc}$. We see that SIDM10~O5 and SIDM30~O7 could reproduce the stellar size and velocity dispersion of Crater~II as SIDM60~O3. Thus there is a degeneracy effect between cross section and orbit. Note that orbits O5 and O7 are not directly motivated by the {\it Gaia} EDR3 measurements and their associated halos would be tidally disrupted after $\sim8~{\rm Gyr}$.

\section{N-body simulations with live stellar particles}
\label{sec:app3}

\begin{figure*}[h]
	      \includegraphics[scale=0.31]{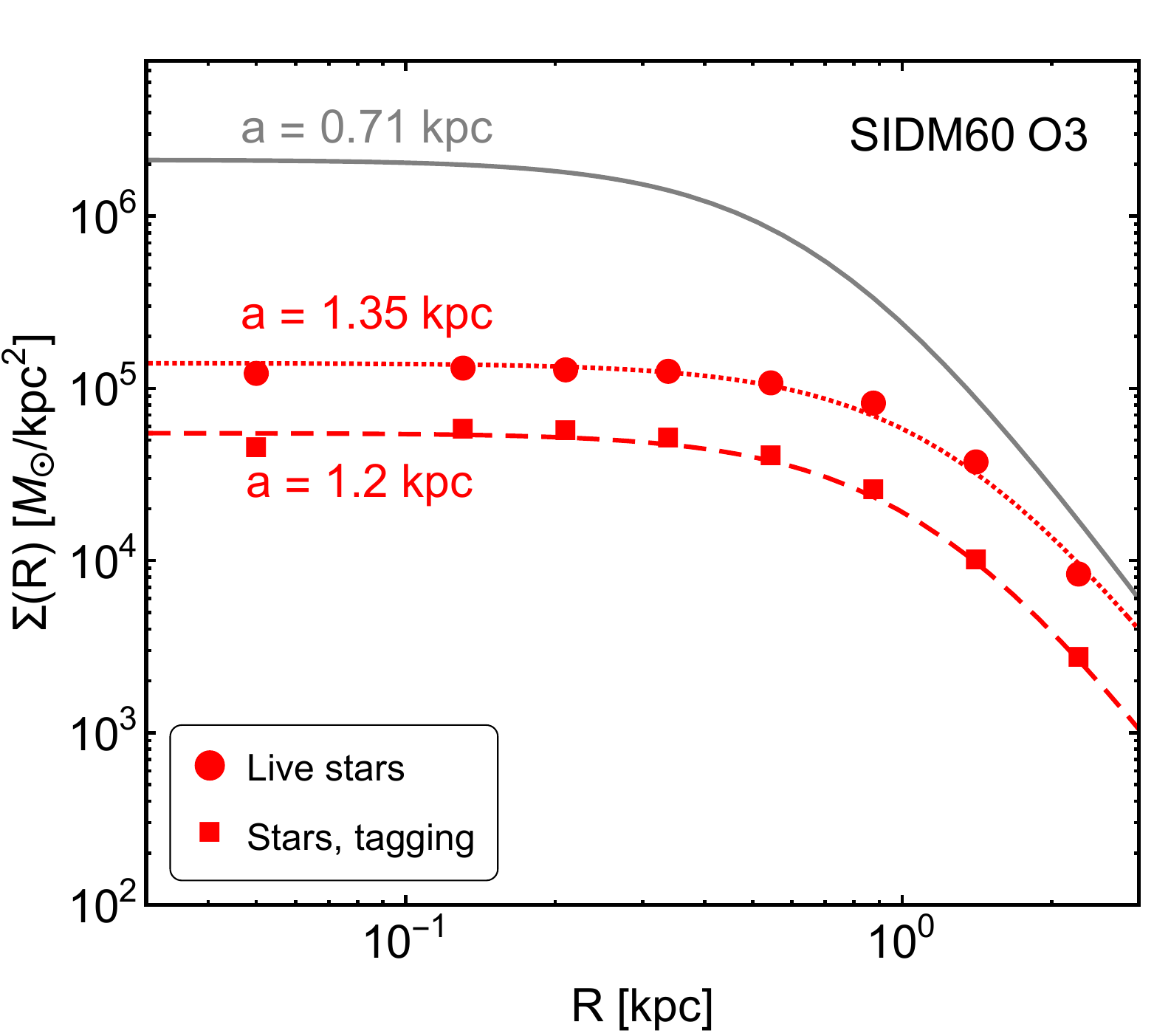}~~~~
		\includegraphics[scale=0.3]{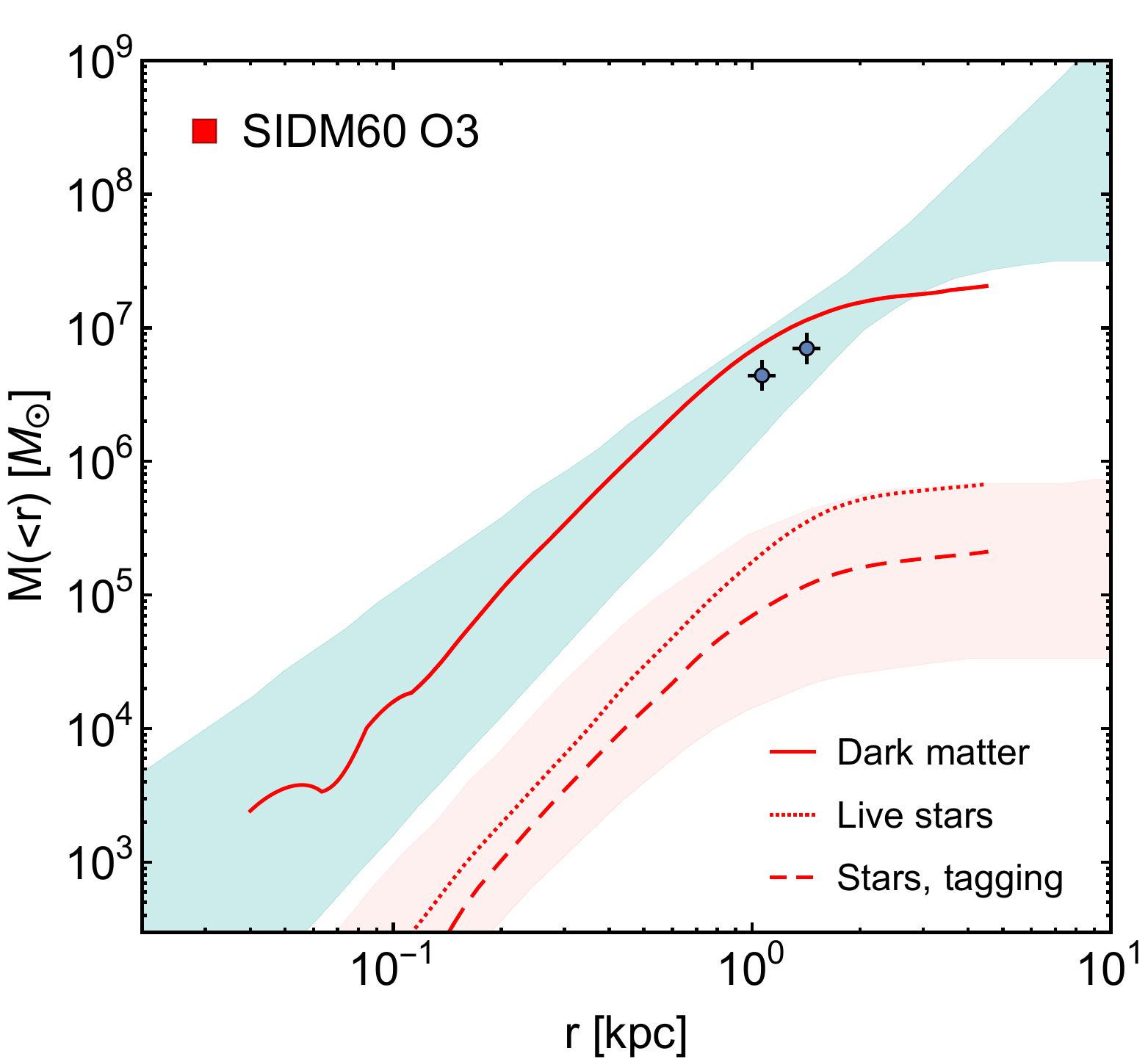}

	\caption{Left panel: Surface density profiles for SIDM60~O3 with {\it live} stellar particles, including the initial condition (solid), the one constructed from live stellar particles at the final snapshot (dot) and its fit to a Plummer profile (dotted), as well as the one constructed from dark matter particles using the tagging method (square) and its fit to a Plummer profile (dashed). Right panel: Final mass profiles for dark matter (solid) and stars constructed from live stellar particles (dotted) and dark matter particles using the tagging method (dashed) for SIDM60~O3. The cyan and pink bands denote the $68\%$ credibility intervals for dark matter and stars, respectively~\citep{caldwell2017crater-CraterII-obs}.}
\label{fig:app3}
\end{figure*}

We further confirm the main results and the tagging method itself by performing additional N-body simulations with {\it live} stellar particles for SIDM60~O3. We assume the initial stellar distribution follows a Plummer profile as
\begin{eqnarray}
\rho(r)=\frac{3M_\star}{4\pi a^3}\left(1+\frac{r^2}{a^2}\right)^{-\frac{5}{2}},
\end{eqnarray}
where $M_\star$ is the total stellar mass and $a$ is the scale radius. We take $M_\star=3.37\times10^6~M_\odot$ and $a=0.71~{\rm kpc}$. For the Plummer profile, the 2D half-light radius is $R_{1/2}=a$, hence the initial value of $R_{1/2}$ is $0.71~{\rm kpc}$, which is comparable with the E1 initial condition for the Einasto profile, i.e., $R_{1/2}\approx2.03r_E=0.81~{\rm kpc}$. We take the Plummer profile for the stars as it is relatively straightforward to implement in the code~\texttt {SpherIC}~\citep{Garrison-Kimmel:2013yys}. For the dark matter halo, we take $\rho_s=1.02\times10^7~M_\odot/{\rm kpc^3}$ and $r_s=2.30~{\rm kpc}$. The concentration of this halo is slightly lower than the one without stars to offset the baryonic influence of live stellar particles. This is a subtle and novel effect of SIDM and we will elaborate it further in a future publication.

Figure~\ref{fig:app3} (left panel) shows the initial (gray) and final (red) surface density profiles for SIDM60~O3 from N-body simulations with live stellar particles. We construct the stellar surface density profile at the final snapshot from stellar particles (dot), fit it with a Plummer profile (dotted), and determine the half-light radius at the final snapshot $R_{1/2}=a\approx1.35~{\rm kpc}$. For comparison, we also use the tagging method to reconstruct the surface density (square), fit it with a Plummer profile (dashed), and find $R_{1/2}\approx1.2~{\rm kpc}$. We see that $R_{1/2}$ at the final snapshot is increased almost by a factor of $2$ compared to the initial condition, due to SIDM core formation. Furthermore, $R_{1/2}$ inferred from the tagging method agrees with that from live stars within $11\%$, although the former underestimates the overall normalization of the surface density by a factor of $3$. We use the tagging method for the purpose of tracking the evolution of the galaxy size, and we can add a normalization factor to match the observed surface density. Furthermore, we have checked that for SIDM60~O3, the size evolution is almost the same whether using Einasto or Plummer profiles in the tagging method.   

We also note that the surface density decreases in the central regions more significantly than in the outer regions. This is because SIDM core formation leads to a shallower density profile and weaker gravitational potential. Thus, the stellar particles more likely migrate from the center towards the outer regions and compensate for the loss there, compared to the CDM limit. 

Figure~\ref{fig:app3} (right panel) shows the final mass profiles for dark matter (solid) and stars constructed from live stars (dotted) and dark matter particles using the tagging method (dashed). We see that the SIDM60 simulation with live stellar particles well reproduces the inferred halo and stellar mass profiles of Crater~II~\citep{caldwell2017crater-CraterII-obs}.

\bibliography{refbib-CraterII} 
\bibliographystyle{aasjournal}

\end{document}